\documentclass[11pt,a4paper]{article}
\usepackage{jheppub}

\usepackage[T1]{fontenc}
\usepackage{amssymb,amsmath}
\usepackage{lmodern}
\usepackage{verbatim,setspace}

\DeclareMathOperator{\R}{Re}
\DeclareMathOperator{\I}{Im}
\newcommand{\ee}{\mathrm{e}}

\newcommand{\Iprod}[2]{\langle {#1}, {#2} \rangle}
\newcommand{\cA}{\mathcal{A}}
\newcommand{\cB}{\mathcal{B}}
\newcommand{\cC}{\mathcal{C}}
\newcommand{\cG}{G}
\newcommand{\cK}{\mathcal{K}}
\newcommand{\cF}{\mathcal{F}}
\newcommand{\cV}{\mathcal{V}}
\def\cH{{\mathcal H}}
\newcommand{\im}{\mathrm{i}}
\newcommand{\nv}{n_\mathrm{v}}
\newcommand{\cN}{\mathcal{N}}
\def\squad{\,\,\,}
\newcommand{\cHz}{ \cH_{ \mbox{\tiny{0}} } }
\newcommand{\ord}[1]{{\scriptscriptstyle (#1)}}

\usepackage{array}
\usepackage{hhline}

\makeatletter
\newcommand{\thickhline}{%
    \noalign {\ifnum 0=`}\fi \hrule height 1.3pt
    \futurelet \reserved@a \@xhline
}
\newcolumntype{"}{@{\hskip\tabcolsep\vrule width 1pt\hskip\tabcolsep}}
\makeatother
\newlength{\arrayrulewidthOriginal}

\title{Static BPS black holes in $U(1)$ gauged supergravity}
\author{Stefanos Katmadas}
\affiliation{Dipartimento di Fisica, Universit\'a di Milano-Bicocca,
I-20126 Milano, Italy
\\ and \\
INFN, sezione di Milano-Bicocca}

\emailAdd{stefanos.katmadas [at] unimib.it}

\abstract{ 
We consider the flow equations for $1/4$-BPS asymptotically AdS$_4$ static black holes in
Fayet-Iliopoulos gauged supergravity, using very special geometry identities to obtain a
simplified form in the most general case. Under mild assumptions on the form of the solution,
we analyse the flow equations and find an explicit solution for arbitrary gauging and charge
vectors within the chosen ansatz. Comparing with the corresponding attractor equations, we find
that the solution is given in terms of exactly the same vector of parameters, implying that all
regular attractors can be extended to full black hole solutions. We present explicit examples
of black hole solutions with all complex scalars and charges allowed by the ansatz turned on,
within the STU model and its truncations.
}

\dedicated{\vspace{2cm} Dedicated to the memory of Christos Katmadas}

\begin{document}
 \maketitle

\section{Introduction}

Supersymmetric backgrounds in supergravity have been a versatile tool for furthering
the understanding of various aspects of string models and supersymmetric theories of
gravity. Despite the rather strong constraints imposed on them, supersymmetric
solutions very often provide a rich subsector of a supergravity theory that is at the
same time simple enough to be studied analytically, granting good control over various
quantities that are otherwise difficult to study. When specified to black holes embedded
in flat spacetime, there is a long list of important results, based on the classification
of supersymmetric solutions in both four and five dimensions
\cite{Behrndt:1997ny, Denef:2000nb, Gauntlett:2002nw, Gauntlett:2004qy}, including
various novel solutions and important insights in the dual string theoretic picture.
Unlike the asymptotically flat solutions, there is much less known about BPS solutions
with AdS asymptotics, as a general classification is missing. The known solutions
include various black hole and black brane solutions that asymptote either to AdS$_4$
or AdS$_5$, that have been obtained by various methods.

In particular, BPS black hole solutions in AdS$_4$
\cite{Cacciatori:2008ek,Klemm:2010mc,Meessen:2012sr}, have been the subject
of considerable recent interest, starting with the first AdS$_4$ solution with a
spherical horizon, obtained in \cite{Cacciatori:2009iz}, which was later expanded upon
with the work of \cite{Dall'Agata:2010gj, Hristov:2010ri, Klemm:2011xw, Barisch:2011ui, Colleoni:2012jq,
Halmagyi:2013sla, Halmagyi:2013qoa, Barisch-Dick:2013xga, Gnecchi:2013mta, Halmagyi:2013uza}.
These extensions include various analytical and numerical solutions, mainly describing
static backgrounds, as well as some notable stationary solutions.
A common feature of all these BPS solutions in AdS$_4$ is the fact that the scalars
are usually restricted to special configurations throughout the spacetime, in order to
render the equations tractable. Indeed, almost all the known solutions are given for
gaugings that allow for vanishing axions at infinity, which are then assumed to vanish
everywhere, so that only half of the BPS flow equations are relevant.

In this paper we consider the extension of these results to the most general case, for
static $1/4$-BPS black holes in Fayet-Iliopoulos (FI) gauged supergravity with
symmetric scalar manifold. Using techniques of very special geometry, we show that the
BPS flow equations can be drastically simplified, so that they involve only one combination
of the scalars, which in addition transforms covariantly under electric/magnetic duality
reparametrisations.

The resulting equations are again nonlinear and therefore more complicated than the
corresponding ones in the ungauged theory \cite{Denef:2000nb}. Nevertheless, they allow
for much better analytic control for general gaugings and scalar configurations. Employing
a general expansion in powers of the radial coordinate in AdS$_4$, we obtain a very
restricted system that precisely corresponds to the standard ansatz of
\cite{Cacciatori:2009iz, Dall'Agata:2010gj, Hristov:2010ri, Gnecchi:2013mta, Halmagyi:2013uza},
but allows for a general flow of the complex scalars. The latter is parametrised in terms of
a single symplectic vector, which is subject to a number of constraints and is determined by
the gauging and the electromagnetic charges.

As already noted in \cite{Dall'Agata:2010gj, Hristov:2010ri, Halmagyi:2013qoa, Gnecchi:2013mta, Halmagyi:2013uza},
the charges are in fact not completely free, but are restricted for a given gauging vector.
In order to systematically study the possible solutions, we recast the attractor equations
in terms of a single real symplectic vector that describes both the values of the scalar
fields and the charges. Upon comparison with the vector parametrising the general
asymptotically AdS$_4$ solution, we find the two vectors to be exactly equivalent, implying
that any regular attractor geometry in the standard branch\footnote{As explained in section
\ref{sec:attr}, there are two branches of attractors, only one of which appears to be
relevant} can be extended to a full black hole geometry.

This paper is organised as follows. In section \ref{sec:attr} we revisit the attractor
geometry in AdS$_4$, recasting the attractor equations in a form that is particularly
well suited for the comparison with the full flow. In addition, this provides a simple example  
of the technique used in the discussion of the full BPS equations. The latter is presented in
section \ref{sec:AdS}, where we recast the BPS equations in a form that can be analysed by
elementary methods without the need of assumptions on the form of the gauging or scalars.
Section \ref{sec:examples} is devoted to a number of explicit examples of black holes with
both electric and magnetic charges turned on, within the STU model and its truncations.
We conclude in section \ref{sec:conc}, where we discuss possible extensions of the results
of this paper. The two appendices give the basic conventions used throughout the paper and
some very useful identities satisfied by the quartic invariant of very special geometry.

\vspace{0.5cm}
{\bf Note added:} The attractor branch mentioned above can be in fact generalised in a
nontrivial way, allowing for more general solutions to the flow equations considered in
this paper. We refer to the notes at the end of sections \ref{sec:attr} and \ref{sec:flow-ana}
for further details.

\section{Revisiting BPS attractors in AdS$_4$}
\label{sec:attr}
In this section we revisit the BPS attractor equations in FI gauged supergravity
\cite{Bellucci:2008cb, Klemm:2010mc, Dall'Agata:2010gj, Hristov:2012nu, Halmagyi:2013qoa},
recasting the symplectic covariant equations in a form that can be easily
connected to the flow equations for the full asymptotically AdS$_4$ black hole
geometries. Our analysis is based only on symplectic vectors in the real basis
and is therefore somewhat complimentary to the corresponding analysis of
\cite{Halmagyi:2013qoa}, which relied on the complex basis.

The starting point is the expressions for the metric and (constant) scalar fields
at the AdS$_2\times$S$^2$ attractor, which take the form
\begin{gather}\label{metric-fin}
  ds^2 = -\ee^{2U_0} r^2\,d t^2  + \ee^{-2U_0} \frac{dr^2}{r^2} 
  + \ee^{2(\psi_0-U_0)}\, \left( d\theta^2 + \sin^2{\theta} d\phi^2 \right)\,,
  \\
2\,\ee^{2\psi_0 - U_0}\I(\ee^{-\im\alpha}\cV) = \,
 \Gamma + \ee^{2(\psi_0-U_0)} \mathrm{J} \cG  \, .
\label{eq:attr-or}
\end{gather}
Here, $\Gamma$ is the vector of charges, which completely fixes the field strengths
in the static case through \eqref{eq:dual-gauge}, while $\cG$ is a symplectic
vector of electric and magnetic FI terms and $\mathrm{J}$ is the scalar dependent
complex structure defined in Appendix \ref{app:conv}. The positive constants
$\ee^{U_0}$ and $\ee^{\psi_0}$ control the radii of AdS$_2$ and S$^2$ respectively
and have been chosen for convenience in connecting with the full flow in later
sections. The phase $\ee^{-\im\alpha}$ is such that the central
charges\footnote{See \eqref{ch-def} for a general definition of central charges.}
of the charge, $Z(\Gamma)$, and the gauging, $Z(\cG)$, satisfy
\begin{equation}\label{eq:RZIW}
- \ee^{2(U_0-\psi_0)}\,\R(\ee^{-\im\alpha} Z(\Gamma)) 
= \I(\ee^{-\im\alpha} Z(\cG)) = \tfrac12\, \ee^{U_0}\,,
\end{equation}
while its complex conjugate is also trivially satisfied, due to the
additional conditions
\begin{gather}\label{eq:RWIZ}
\I(\ee^{-\im\alpha} Z(\Gamma)) = \R(\ee^{-\im\alpha} Z(\cG)) =0 \,,
\\
\label{eq:att-Dir}
\Iprod{\cG}{\Gamma}=-1\,.
\end{gather}
The latter originate from the fact that the BPS equations relate these quantities
to the (vanishing) K\"ahler connection and the spin connection respectively.

In order to characterise the solution to the above equations, we start by
parametrising the symplectic section in a convenient way in terms of a
vector, $\cB$, as
\begin{align} 
 2\,\ee^{\psi_0-U_0}\I(\ee^{-\im\alpha}\cV)= \cB \,,
\end{align}
and solve for the second of \eqref{eq:RZIW}, so that $\psi_0$ and the
section become
\begin{gather}\label{eq:attr-scal-r0}
 2\,\ee^{-U_0}\I(\ee^{-\im\alpha}\cV)= \Iprod{G}{\cB}^{-1}\cB \,,
 \\
 \ee^{\psi_0} = \Iprod{G}{\cB}\,.
 \label{eq:psi-attr}
\end{gather}
In terms of $\cB$, the scale factor and the real part of the section are given by
\begin{equation}\label{eq:real-sec}
\ee^{-4U_0} = \Iprod{G}{\cB}^{-4} I_4(\cB)\,,
\qquad
2\,\ee^{U_0}\R(\ee^{-\im\alpha}\cV)= \tfrac12\,\Iprod{G}{\cB}\,I_4(\cB)^{-1}I^\prime_4(\cB) \,.
\end{equation}
The next step is to insert this vector in the \eqref{eq:attr-or} to obtain an
equation for the vector $\cB$, which requires an expression for the action of
the scalar dependent operator $\mathrm{J}$ on the vector of gauging, $\cG$, in
terms of $\cB$. As explained in Appendix \ref{app:I4}, this can be done using
equation \eqref{I4toJ}, which when evaluated for $\cG$, reads
\begin{equation}
 \tfrac12\, I^\prime_4(\cB, \cB, \cG )
    = 2\,\Iprod{\cG}{\cB}\,\cB
    -2\,\Iprod{G}{\cB}^{2}\ee^{-2U_0}\mathrm{J}\,\cG \,,
    \label{JG}
\end{equation}
where we used the definition \eqref{eq:attr-scal-r0} and \eqref{eq:RWIZ}. Inserting
this in \eqref{eq:attr-or} we obtain
\begin{equation}\label{eq:Gam-BB}
 \tfrac14\, I^\prime_4(\cB, \cB, \cG ) = \Gamma\,,
\end{equation}
which can be solved to find the vector $\cB$ in terms of the gauging and the charge
(see section \ref{sec:examples} for example solutions).
The scalar fields and metric components are then given by \eqref{eq:attr-scal-r0},
\eqref{eq:real-sec} and \eqref{eq:psi-attr} respectively. Note that all BPS equations
are satisfied except \eqref{eq:RWIZ}-\eqref{eq:att-Dir}, which lead to the
constraints
\begin{equation}\label{eq:attr-B}
\Iprod{\cB}{\Gamma}=I_4(\cB,\cB,\cB,\cG)=0\,, \qquad \tfrac14\, I_4(\cB, \cB, \cG, \cG ) = -1\,.
\end{equation}
Note that the two equations in \eqref{eq:RWIZ} have become linearly dependent due to
\eqref{eq:Gam-BB}, in line with the fact that only one of them fixes a physical phase,
the second one being the unphysical phase $\alpha$. Therefore, \eqref{eq:attr-B} reduce
the $2\,\nv+2$ components of $\cB$ to $2\,\nv$ independent components.
Finally, the black hole entropy, computed by the horizon area is given by the
simple expression
\begin{equation}\label{eq:entr-form}
\mathcal{E}=\pi\,\sqrt{I_4(\cB)}\,.
\end{equation}
Note that this expression is reminiscent of the entropy formula for  asymptotically flat
extremal solutions, which is obtained from \eqref{eq:entr-form} by replacing $\cB$ by the
charge vector $\Gamma$.

For future reference, we briefly discuss a more direct formula for the entropy, involving
only charges and gaugings, obtained in \cite{Halmagyi:2013qoa} by rearranging
\eqref{eq:attr-or}-\eqref{eq:RWIZ} into the following relation
\begin{equation}\label{eq:hol-attr}
 \Gamma + \mathrm{i}\,\ee^{2(\psi_0-U_0)}\cG 
 = \mathrm{i}\,\bar{Z}(\Gamma)\,\cV - \mathrm{i}\,g^{\bar \imath j} \bar{Z}_{\bar \imath}(\Gamma)\, D_j \cV\,.
\end{equation}
The last relation is purely holomorphic in the the symplectic section and its derivatives, so
that one directly obtains from \eqref{I4-def} that
\begin{equation}
 I^\prime_4(\Gamma + \mathrm{i}\,\ee^{2(\psi_0-U_0)}\cG)=\tfrac16\,c_{ijk} \bar Z^i \bar Z^j \bar Z^k\,\cV\,,
 \qquad
 I_4(\Gamma + \mathrm{i}\,\ee^{2(\psi_0-U_0)}\cG)=0\,.
 \label{eq:I4CH0}
\end{equation}
These are solved by the above expressions in terms of $\cB$, but the second can be easily
solved for the radius of the sphere as\footnote{Note that there exists another branch of
solutions, for which $\ee^{4(\psi_0-U_0)}\sim I_4(\cG,\Gamma,\Gamma,\Gamma)/I_4(\cG,\cG,\cG,\Gamma)$, see
\cite{Halmagyi:2013qoa} for details. Here, we display only the branch that arises from the
BPS equations in the bulk, as derived in the next section. It is an interesting problem to
find physically relevant situations where this second branch arises.}
\begin{gather}
 \ee^{4(\psi_0-U_0)} = I_4(\cB) = 
 \frac{1}{2\,I_4(\cG)}\,\left(\frac{1}{4}\,I_4(\cG,\cG,\Gamma,\Gamma)
\pm \sqrt{\frac{1}{16}\,I_4(\cG,\cG,\Gamma,\Gamma)^2- 4\,I_4(\cG)I_4(\Gamma)}\right)\,,
\label{eq:entr-exp}
\\
I_4(\cG,\cG,\cG,\Gamma)=I_4(\cG,\Gamma,\Gamma,\Gamma)=0
\label{eq:bulk-con}
\,,
\end{gather}
where only the plus sign leads to a positive radius for all examples we discuss in section \ref{sec:examples}. 
The first of these arises from the real part of \eqref{eq:I4CH0} and provides the required
entropy formula, while \eqref{eq:bulk-con} arises from the imaginary part and represents a
constraint on the charges, since the two conditions are linearly related upon using the result
\eqref{eq:Gam-BB} above, together with the identities presented in appendix \ref{app:I4}. 

As will be shown in the next section, the flow to asymptotic AdS$_4$ imposes \eqref{eq:bulk-con},
so that this additional condition is necessary for extending the attractor to a full black hole
geometry. In fact, the treatment of the next section allows for the full solution to be
constructed if the attractor solution for the scalars is known, in exactly the same way as for
asymptotically flat black holes. 

Obtaining the attractor solution involves solving \eqref{eq:Gam-BB}, which is a complicated
task in general, since it corresponds to a $2(\nv+1)$ dimensional system of quadratic equations,
on which three constraints have to be imposed. We have two constraints from \eqref{eq:attr-B}
and an additional one from \eqref{eq:bulk-con}, reducing $\cB$, and therefore also $\Gamma$, to
$2\,\nv-1$ independent components. For restricted cases, it is possible to solve this system for
all symmetric models, see \cite{Gnecchi:2013mta} for a closely related
computation\footnote{Note that the vector $\cB$ used here is related to the one used in
\cite{Gnecchi:2013mta} by a shift of the radial coordinate, cf. \eqref{eq:B-shift} and the
relevant discussion below}. In section \ref{sec:examples} we present explicit solutions to the
STU model and its truncations. The derivation of a general solution for $\cB$ in terms of the
gauging and the charge vector falls outside the scope of this work.

\vspace{0.5cm}
{\bf Note added:} The condition \eqref{eq:bulk-con} is in fact a special solution to the
system of attractor constraints \cite{Halmagyi:2013qoa}. While this is not obvious
from \eqref{eq:attr-B}, the possibility of a linear dependence between the resulting equations
exists, but is not taken into account in the counting of independent parameters above. Such a linear
dependence was in fact shown to be true in \cite{Halmagyi:2014qza}, leading to $2\,\nv$ free
parameters. We refer to that work for more details.

\section{Asymptotically \protect{AdS$_4$} BPS black holes}
\label{sec:AdS}

In this section, we consider the full asymptotically AdS$_4$ flow for $1/4-$BPS static black holes,
generalising the analysis of the attractor geometry presented in the previous section. This is based
on the static flow equations as derived in \cite{Dall'Agata:2010gj}, whose conventions we follow up
to some changes in naming.

\subsection{Analysis of the flow equations}\label{sec:flow-ana}
For the class of solutions we are interested in, the appropriate ansatz for a static metric is
\begin{gather}\label{4dmetric}
  ds^2 = -\ee^{2U} d t^2  + \ee^{-2U} \left( dr^2 + \ee^{2\psi} d\theta^2 + \ee^{2\psi} \sin^2{\theta} d\phi^2 \right)\,,
\end{gather}
which allows for a non-flat three dimensional base. In these variables, the boundary conditions for
the metric fields at infinity are given by
\begin{equation}\label{eq:asymp}
 \ee^\psi = I_4(\cG)^{1/4}\,r^2 + \mathcal{O}(r)\,, 
 \qquad
 \ee^{U} = I_4(\cG)^{1/4}\,r + \mathcal{O}(r^0) \,,
\end{equation}
where we used the requirement that $R_{AdS}=I_4(\cG)^{-1/4}$ is the radius of the asymptotic AdS$_4$.
The BPS flow equations are \cite{Dall'Agata:2010gj}:
\begin{gather}
2e^{2 \psi}\left(e^{-U}{\mathrm{Im}}(e^{-i\alpha}{\cal V})\right)'
+e^{2(\psi-U)}\, \mathrm{J} \cG
+4e^{2 \psi-U}(Q_r + \alpha'){\mathrm{Re}}\left(e^{-i\alpha}{\cal V}\right)
+\Gamma = 0, \label{E0-ads}\\
Q_r + \alpha'= -2e^{-U}\,\,{\rm Re}(e^{-i \alpha} Z\left(\cG)\right) \label{Kah-conn}\\
\psi'=2\,e^{-U}\,{\rm Im}\left(e^{-i \alpha} Z(\cG)\right), \label{eq:psi-AdS}\,,
\end{gather}
where the charge and gauging are required to satisfy the quantisation
condition
\begin{equation}\label{Dir-quant0}
\Iprod{\cG}{\Gamma}=-1\,.
\end{equation}
In writing these equations, we used the definition of the central charge functions in
\eqref{ch-def}, which depend on the scalar fields and are defined everywhere in spacetime,
rather than the central charge defined at infinity.

We now proceed to recast these BPS equations in a simpler form, so that they depend on
the scalars only through the imaginary part of the section.
Similar to the analysis of the attractor, we use again the identity \eqref{I4toJ} to
express the action of the complex structure, $\mathrm{J}$, on the gauging as
\begin{align}
 \tfrac14\, I^\prime_4\left(2 \I(e^{-i\alpha}{\cal V}),\, 2 \I(e^{-i\alpha}{\cal V}),\,\cG\right)
 =&\, - \,\mathrm{J}\,\cG
      + 4\,\I(e^{-i\alpha} Z(\cG))\,\I(e^{-i\alpha}{\cal V})
    \nonumber\\
  &\, +8\,\R(e^{-i\alpha} Z(\cG))\,\R(e^{-i\alpha}{\cal V}) \,.
    \label{JG-gen}
\end{align}
Using \eqref{Kah-conn} to rewrite the flow equation for the section as
\begin{align}
2\ee^{2 \psi}\left(\ee^{-U}{\mathrm{Im}}(\ee^{-i\alpha}{\cal V})\right)'
&
+e^{2(\psi-U)}\,\mathrm{J} \cG
-8\,\ee^{2(\psi-U)}\,{\rm Re}(e^{-i \alpha} Z(\cG))\,
 {\mathrm{Re}}(\ee^{-i\alpha}{\cal V}) +\Gamma
=0
\,,
\end{align}
and comparing with \eqref{JG-gen}, we find the following simplified flow equation,
where the scalar fields appear solely through the symplectic section
\begin{align} 
2\ee^{2 \psi}\left(\ee^{-U}{\mathrm{Im}}(\ee^{-i\alpha}{\cal V})\right)'
&\,+ 4\,e^{2(\psi-U)}\,\I(e^{-i\alpha} Z(\cG))\,\I(e^{-i\alpha}{\cal V})
\nonumber\\
&\,-\tfrac14\,e^{2(\psi-U)}\, I^\prime_4\left(2 \I(e^{-i\alpha}{\cal V}),\, 2 \I(e^{-i\alpha}{\cal V}),\,\cG\right)
+\Gamma
=0 \,.
\end{align}
The first two terms can be combined upon use of \eqref{eq:psi-AdS} to obtain
\begin{align}\label{E-simple}
2\ee^{\psi}\left(\ee^{\psi-U}{\mathrm{Im}}(\ee^{-i\alpha}{\cal V})\right)'
&\,-\tfrac14\,e^{2(\psi-U)}\, I^\prime_4\left(2 \I(e^{-i\alpha}{\cal V}),\, 2 \I(e^{-i\alpha}{\cal V}),\,\cG\right)
+\Gamma
=0 \,,
\end{align}
which, together with \eqref{eq:psi-AdS} form a system of first order equations for the
quantities $\ee^\psi$ and 
\begin{equation}
 2\,\ee^{\psi-U}\I(\ee^{-\im\alpha}\cV) \equiv \cH\,,
\end{equation}
alone, as
\begin{gather}
 (\ee^\psi)'= \Iprod{\cG}{\cH}\,, 
 \label{eq:psi-prime}
 \\ 
 \ee^\psi \cH' -\tfrac14\,I^\prime_4(\cH,\cH,\cG) +\Gamma =0 \,.
 \label{eq:flow-fin}
\end{gather}

Despite the non-linearity of \eqref{eq:flow-fin}, it is simple to find solutions to this system,
using as input the fact that $\ee^\psi$ is a regular function which behaves $\sim\! r^2$
asymptotically and has a single zero at the horizon. Since the first of \eqref{eq:psi-prime} is
linear, the vector $\cH$  must be at most linear in the radial coordinate. One may then consider
an expansion consistent with the standard asymptotic expansion in AdS$_4$, as
\begin{equation}\label{eq:H-ans}
 \cH = \cA \, r + \cB + \sum_{n \geq 1} \cC_n \, \frac{1}{r^{n}}\,,
\end{equation}
where $\cA$, $\cB$ and the $\cC_n$ for $n\geq 1$ are constant vectors. For simplicity, we set
all the vectors $\cC_n=0$ for the moment and perform the analysis for the linear terms only,
postponing the justification of this truncation at the end of this subsection.
The equation for $\ee^\psi$ can now be easily integrated as
\begin{align}
 \Iprod{G}{\cH}=&\, 2\,r\,I_4(\cG)^{1/4} + \Iprod{G}{\cB} \,, 
 \nonumber\\ 
 \ee^\psi=&\, I_4(\cG)^{1/4}\,r^2 + \Iprod{G}{\cB}\, r + c \,,
 \label{psi-sol}
\end{align}
where $c$ is an arbitrary constant and we used \eqref{eq:asymp} to determine the coefficient
of the leading term in $\ee^\psi$. 

We now consider each of the three types of terms arising in \eqref{eq:flow-fin} upon using
\eqref{psi-sol}, namely constant, linear and quadratic in the radial coordinate, $r$.
The quadratic terms lead to
\begin{align}\label{A-sol}
I_4(\cG)^{1/4}\,\cA
-\tfrac1{4}\, I^{'}_{4}(\cA,\cA,\cG)  =0
\quad \Rightarrow \quad
\cA = \tfrac12\,I_4(\cG)^{-3/4}\,I^{'}_{4}(\cG)
\,,
\end{align}
which is the natural combination homogeneous in $\cG$ and agrees with the result of \cite{Gnecchi:2013mta}.
The terms linear in $r$ lead to
\begin{align}\label{AB-id}
 \Iprod{\cG}{\cB}\,\cA-\tfrac1{2}\,I^{'}_{4}(\cA,\cB,\cG) =0
 \quad\Rightarrow\quad
\Iprod{\cA}{\cB}=0\,,
\end{align}
where we used the explicit form of $\cA$ in \eqref{A-sol} and \eqref{eq:proj} to obtain
the second equality. Finally, the constant part of
\eqref{eq:flow-fin} reads
\begin{align}\label{B-sol}
c\,\cA
-\frac1{4}\, I^{'}_{4}(\cB,\cB,\cG)  +\Gamma = 0 \,,
\end{align}
which determines $\cB$ and $c$ in terms of the charges and the gauging. Taking the inner
product with $\cG$, we obtain
\begin{align}
 c=\frac{1}{2}\,I_4(\cG)^{-1/4}\left( -\Iprod{\cG}{\Gamma} + \frac1{4}\, I_{4}(\cB,\cB,\cG,\cG) \right)\,,
\end{align}
while the inner product of \eqref{B-sol} with $\cA$ and $\cB$ leads to
\begin{equation}\label{eq:ABGam}
\Iprod{\cA}{\Gamma}=0\,, \qquad
\Iprod{\cB}{\Gamma}=\frac1{4}\, I_{4}(\cB,\cB,\cB,\cG)\,,
\end{equation}
respectively, where we used \eqref{AB-id} to obtain the first result.

The final equation to be imposed is the condition \eqref{Kah-conn}, so we compute
each term individually
\begin{align}
Q_r + \alpha'=&\, -\tfrac12\,\ee^{2(U-\psi)}\Iprod{\cA}{\cB}= 0\,,
\\
2 e^{U}\,\,{\rm Re}(e^{-i \alpha} W) =&\, \tfrac12\,\ee^\psi\,I_4(\cH)^{-1}\Iprod{\cG}{I_4^\prime(\cH)}
\nonumber\\
=&\,\tfrac12\,\ee^\psi\,I_4(\cH)^{-1}\left(\Iprod{\cG}{\cB}\Iprod{\cB}{\cA} r + \tfrac16\,I_4(\cG,\cB,\cB,\cB)\right)
\nonumber\\
=&\tfrac1{12}\,\ee^\psi\,I_4(\cH)^{-1}I_4(\cB,\cB,\cB,\cG) \,,
\end{align}
where \eqref{eq:real-sec} was used for the real part of the section and \eqref{AB-id} was
used repeatedly to simplify the result. We therefore find that \eqref{Kah-conn} is
satisfied if we set
\begin{equation}\label{eq:BGam}
 I_{4}(\cB,\cB,\cB,\cG)=\Iprod{\cB}{\Gamma}=0\,,
\end{equation}
where the second equality follows from \eqref{eq:ABGam} above. This concludes our
analysis of the BPS equations.

Finally, we return to the more general ansatz in \eqref{eq:H-ans} and give some evidence for the
exclusion of any terms of negative power in $r$ in that expansion. We thus consider a general
vector $\cH$ and analyse the horizon behaviour of the flow equation \eqref{eq:flow-fin} and its
derivatives, corresponding to a Taylor expansion of the type
\begin{equation}
 \cH = \sum \frac1{n!}\,\cHz^{\ord{n}} (r-r_{\mbox{\tiny{0}}})^n 
     = \cHz + \cHz^{\ord{1}} (r-r_{\mbox{\tiny{0}}}) + \tfrac12\,\cHz^{\ord{2}} (r-r_{\mbox{\tiny{0}}})^2 
        + \tfrac16\,\cHz^{\ord{3}} (r-r_{\mbox{\tiny{0}}})^3 + \dots \,,
\end{equation}
where we denote horizon values by the $\mbox{\small{0}}$ subscript and $\cHz^{\ord{n}}$
is the $n$-th derivative at the horizon, which we take to be at $r=r_{\mbox{\tiny{0}}}$.
The latter is by definition the solution of $\ee^\psi=0$, where \eqref{eq:flow-fin}
becomes identical to \eqref{eq:Gam-BB}, as
\begin{equation}\label{eq:Gam-HH}
 \Gamma = \tfrac14\,I^\prime_4(\cHz,\cHz,\cG)\,.
\end{equation}
Similarly, taking the
derivative of \eqref{eq:flow-fin} and evaluating at the horizon, one finds the eigenvalue equation
\begin{equation}\label{eq:eigen1}
 \Iprod{\cG}{\cHz}\, \cHz^{\ord{1}} - \tfrac12\, I_4^\prime(\cHz^{\ord{1}},\cHz,\cG) =0
 \qquad \Rightarrow\qquad
 \cHz^{\ord{1}} = \cA \,.
\end{equation}
where $\cA$ is as in \eqref{A-sol}, which matches the derivative of $\cH$ at infinity. In order to
obtain this result, we used \eqref{eq:proj} and the requirement $\Iprod{\cHz}{\cHz^{\ord{1}}}=0$,
which follows from \eqref{Kah-conn} at the horizon. Continuing with the higher derivatives of
\eqref{eq:flow-fin} evaluated at the horizon, one finds an eigenvalue equation involving the same
operator as in \eqref{eq:eigen1}, for the various derivatives of $\cH$. For example, the second
derivative leads to
\begin{equation}\label{eq:eigen2}
 2\,\Iprod{\cG}{\cHz}\, \cHz^{\ord{2}} - \tfrac12\, I_4^\prime(\cHz^{\ord{2}},\cHz,\cG) =0\,,
\end{equation}
where we used \eqref{eq:eigen1} to simplify the result, while the third derivative is given by
the eigenvalue equation
\begin{gather}
 3\,\Iprod{\cG}{\cHz}\, \cK^{\ord{3}} - \tfrac12\, I_4^\prime(\cK^{\ord{3}},\cHz,\cG) =0\,,
 \nonumber\\
 \cK^{\ord{3}} = \cHz^{\ord{3}} 
                + 3\, \frac{\Iprod{\cG}{\cHz^{\ord{1}}}}{\Iprod{\cG}{\cHz}}\,\cHz^{\ord{2}}
                -  \frac{\Iprod{\cG}{\cHz^{\ord{2}}}}{4\,\Iprod{\cG}{\cHz}}\,\cHz^{\ord{1}}\,.
 \label{eq:eigen3}
\end{gather}
Therefore, \eqref{eq:eigen2}-\eqref{eq:eigen3} demand the existence of two eigenvalues for the
matrix $I_4(\cHz, \cG)$ that are twice and three times as large as the one in \eqref{eq:eigen1}
respectively. This is not true in general, but leads to constraints on the vector $\cHz$
and therefore the charges through \eqref{eq:Gam-HH}. In particular, since the matrix involved
is symplectic, the corresponding negative eigen values must also be present, so that there
are four conditions resulting from \eqref{eq:eigen2}-\eqref{eq:eigen3}. By consistency, one
further obtains the three conditions
\begin{equation}
\Iprod{ \cHz^{\ord{1}} }{ \cHz^{\ord{2}} } = \Iprod{ \cHz^{\ord{1}} }{ \cHz^{\ord{3}} } 
 = \Iprod{ \cHz^{\ord{3}} }{ \cHz^{\ord{2}} }=0 \,, 
\end{equation}
which further constrain the allowed eigenvectors.

Using exactly the same procedure, one can show that this pattern continues to higher orders, so that
the same operator in \eqref{eq:eigen1}-\eqref{eq:eigen3} must have integer spaced eigenvalues, thus
constraining the charges further. We conclude that adding nonlinear terms to $\cH$, as the $\cC_n$ in
\eqref{eq:H-ans}, is not allowed in general, as that would turn on an arbitrary number of derivatives
of $\cH$ at the horizon that cannot be accommodated by the finite dimensional matrix above. One may
still hope to find a solution for restricted charges, by arranging that the various nonlinear
terms cancel each other at the horizon. Moreover, a more general expansion at the horizon still
remains to be done in principle, allowing for non-integer powers of the radial variable. However,
nonlinear terms in $\cH$ are inconsistent with the shift invariance of the equations along $r$
discussed in section \ref{sec:rep-rad} below, as well as with all known asymptotically AdS$_4$ black
hole solutions, both extremal and non-extremal alike, in which these terms do not appear, see e.g. \cite{Chow:2013gba,Gnecchi:2013mja} for an overview of the relevant ansatze.

\vspace{0.5cm}
{\bf Note added:} In all the above, the condition \eqref{AB-id}, justified by the attractor analog
in \eqref{eq:bulk-con}, was used as an essential linearly independent constraint on the flow, so
that any solutions not satisfying this assumption can evade the above arguments. See the note at
the end of section \ref{sec:attr} on the existence of such solutions.

\subsection{Reparametrisation of the radius}
\label{sec:rep-rad}
It is interesting to note a redundancy of the above equations, that will prove
useful in the discussion of explicit examples. It is simple to check using
\eqref{B-sol}, combined with \eqref{A-sol}-\eqref{AB-id}, that the shift
\begin{equation}\label{eq:B-shift}
\cB \rightarrow \cB + b \,\cA\,,
\end{equation}
for any constant, $b$, can be reabsorbed by a shift of the radial coordinate,
$r$, both from the scalars and the metric. For example, shifts in
$\Iprod{G}{\cB}$ and $c$, which are linear and quadratic in $b$ respectively,
are such that \eqref{psi-sol} becomes
\begin{equation}
\ee^\psi= I_4(\cG)^{1/4}\,(r + b)^2 + \Iprod{G}{\cB}\,(r + b)+ c\,.
\end{equation}
Turning the argument around, one can always shift the radial variable so
as to arrange that \eqref{eq:B-shift} leads to
\begin{equation}
\Iprod{\cG}{\cB}=0\,,
\end{equation}
holds, which is the choice usually taken in the literature.

Alternatively, one may use this freedom to impose $c=0$. In this case,
we find
\begin{equation}
\ee^\psi= \left(I_4(\cG)^{1/4}\,r + \Iprod{G}{\cB}\right)\, r\,,
\end{equation}
where we assume without loss of generality that
\begin{equation}
 \Iprod{G}{\cB} >0\,,
\end{equation}
since one may always shift the nonzero root of $\ee^\psi$ to $\pm\Iprod{G}{\cB}$.
In contradistinction with the previous choice, this inner product is not allowed
to vanish in order to have a well defined horizon.
Setting $c=0$ leads to a simplification of \eqref{B-sol}, but most importantly it
makes the horizon limit clear, as it is always located at $r=0$ by construction.
We will use this second choice for the remainder of this paper, for simplicity.

\subsection{Summary of static AdS$_4$ BPS solutions and the attractor limit}
\label{sec:summ}
For the convenience of the reader we provide a summary of the general 
duality covariant equations describing static AdS$_4$ 1/4-BPS solutions.
The metric is given by
\begin{gather}\label{metric-attr}
  ds^2 = -\ee^{2U}d t^2  + \ee^{-2U} \left( dr^2 + \ee^{2\psi} d\theta^2 + \ee^{2\psi} \sin^2{\theta} d\phi^2 \right)\,,
  \nonumber\\
\ee^\psi= \left(I_4(\cG)^{1/4}\,r + \Iprod{G}{\cB}\right)\, r\,.
\end{gather}
The scale factor $\ee^{2U}$ and the scalar fields are given by
\begin{align}\label{eq:scal-fin}
 2\,\ee^{-U}\I(\ee^{-\im\alpha}\cV)=
   \ee^{-\psi} \left( \cA\,r + \cB \right) \,,
   \qquad
\cA = \frac12\,I_4(\cG)^{-3/4}\,I^{'}_{4}(\cG)\,,
\end{align}
where we note that the asymptotic value of both are completely fixed by
the gaugings, through the vector, $\cA$, as expected. The attractor flow
is governed by the vector, $\cB$, which is fixed in terms of the charge
through 
\begin{align}\label{B-sol-fin}
\frac1{4}\, I^{'}_{4}(\cB,\cB,\cG) = \Gamma \,.
\end{align}
Note that there are some conditions on both $\cB$ and $\Gamma$ that arise
both from supersymmetry and from imposing consistency of the above equations
and read
\begin{gather}
\Iprod{\cA}{\Gamma}=\Iprod{\cA}{\cB}=\Iprod{\Gamma}{\cB}=0\,,
\label{A-comm}
\\
\Iprod{\cG}{\Gamma} = -\kappa\,,
\label{Dir-quant}
\\
\Iprod{\cG}{\cB}>0\,.
\label{BG-bound}
\end{gather}
Here, we have introduced the constant, $\kappa$, generalising \eqref{Dir-quant0} to allow for
solutions with different horizon geometry \cite{Cacciatori:2009iz, Dall'Agata:2010gj}. Setting
$\kappa=1$ leads to black holes with a spherical horizon, whereas $\kappa=-1$ or $\kappa=0$
lead to black holes with a hyperbolic or flat horizons respectively.

At this point it is instructive to count the number of parameters allowed for
the solutions above. Taking the gaugings to be arbitrary, \eqref{A-comm}
impose two conditions on each of $\Gamma$ and $\cB$, reducing them to $2\,\nv$
components each. Note that this is consistent with \eqref{B-sol-fin}, which now
also contains the same number of independent components. The constraint
\eqref{Dir-quant} further reduces the number of free charges to $2\,\nv-1$, while
it implies, through \eqref{B-sol-fin}, that
\begin{equation}\label{eq:B-constr}
\frac1{4}\, I_{4}(\cB,\cB,\cG,\cG) = -1\,,
\end{equation}
which can be viewed as a normalisation condition on the solutions of \eqref{B-sol-fin}.

Comparing with asymptotically flat black holes, it is useful to note that \eqref{A-comm}
still holds in that case\footnote{In the asymptotically flat case, the vector $\cB$
is not relevant.} and expresses the requirement of vanishing NUT charge. Exactly as in
the above analysis, the vector, $\cA$, is again parametrising the asymptotic scalars.
The latter are arbitrary in the ungauged theory, so $\cA$ is also a priori arbitrary
and is only constrained by \eqref{A-comm}. In the gauged theory however, the asymptotic
scalars are completely fixed by the gauging through \eqref{eq:scal-fin}, so that
\eqref{A-comm} instead becomes a constraint on the allowed charge.

Finally, we connect with the attractor analysis of section \ref{sec:attr}, by
taking the near horizon limit. In the chosen coordinates this is located at $r=0$,
so that one only needs to expand the above expressions for small $r$. The metric
scale factors take the form
\begin{equation}
\ee^\psi \rightarrow \Iprod{G}{\cB} \, r \equiv \ee^{\psi_0} r\,,
\qquad
\ee^U \rightarrow \Iprod{G}{\cB}\,I_4(\cB)^{-1/4} \, r \equiv \ee^{U_0} r\,,
\end{equation}
where we defined the constants $\ee^{\psi_0}$ and $\ee^{U_0}$. Similarly, the near
horizon limit of the scalars leads to
\begin{equation}
2\,\ee^{-U_0}\I(\ee^{-\im\alpha}\cV) = \Iprod{G}{\cB}^{-1} \cB \,.
\end{equation}
These expressions are identical to \eqref{eq:attr-scal-r0}, \eqref{eq:psi-attr} and
\eqref{eq:real-sec} derived at the attractor. Similarly, \eqref{B-sol-fin} matches
with \eqref{eq:Gam-BB} and \eqref{A-comm}-\eqref{Dir-quant} impose the constraints
\eqref{eq:attr-B} and \eqref{eq:bulk-con}. We conclude that any attractor solution
satisfying to the conditions presented in section \ref{sec:attr} can be extended to
an asymptotically AdS$_4$ black hole.

\section{Explicit examples}
\label{sec:examples}
In this section we consider some explicit solutions for three cubic models,
namely the STU model and its two straightforward truncations, the two-modulus
$st^2$ model and the one-modulus $t^3$ model, which are the benchmark examples
for all the symmetric models in the infinite $SO(2,n+1)$ series. While our discussion
is by no means exhaustive, we illustrate the salient features of BPS solutions,
focusing on recovering the largest possible number of parameters. In particular, we
recast the magnetic solutions of \cite{Cacciatori:2009iz, Hristov:2010ri, Dall'Agata:2010gj}
in the conventions of this paper, so that the horizon is at a fixed locus $r=0$.
In addition, we display explicit expressions for the dyonic solutions constructed recently
in \cite{Halmagyi:2013uza} by duality rotations on the magnetic solutions. While all
the solutions for the STU model presented in the papers above and in the present section,
should be limits of the general solution given recently in \cite{Chow:2013gba}, this
connection is difficult to implement explicitly.

The STU model is defined by the prepotential
\begin{equation}\label{STU}
 F= \frac{X^1X^2X^3}{X^0}\,,
\end{equation}
and contains three physical complex scalar fields that appear completely symmetrically,
each coming from a vector multiplet.
In all examples in this section, we will choose the gauging vector, $G$ to be such
that the axions vanish on the AdS$_4$ vacuum at infinity. In particular, we take
\begin{equation}\label{eq:G-STU}
\cG=\left( g^0, \squad 0, \squad 0, \squad 0, \squad g_1,\squad g_2, \squad g_3, \squad 0 \right){}^{\!T}\,,
\end{equation}
for the STU model, while we reduce this vector by taking $g_3=g_2$ for the $st^2$
model and $g_3=g_2=g_1$ for the $t^3$ model. Note that \eqref{eq:G-STU} can be easily
rotated to an electric frame, at the expense of making \eqref{STU} more complicated.
Given this vector, the radius of AdS$_4$ and the asymptotic scalars are controlled by
the vector $\cA$ in \eqref{eq:scal-fin}, which reads
\begin{gather}\label{eq:A-STU}
\cA = \tfrac12\, I_4(G)^{-3/4} I^\prime_4(G) = \tfrac1{\sqrt{2}}\,(-g^0 g_1 g_2 g_3)^{1/4}
\left( 0, \squad 1/g_1, \squad 1/g_2, \squad 1/g_3, \squad 0,\squad 0, \squad 0, \squad -1/g_0 \right){}^{\!T}\,,
\end{gather}
so that we find 
\begin{gather}\label{eq:vac-STU}
R_{AdS} = I_4(G)^{-1/4}= (-4 \, g^0 g_1 g_2 g_3)^{-1/4}\,, 
\qquad
t^i\bigr|_{\infty} =\mathrm{i}\,\sqrt{-\frac{g_1g_2g_3}{g^0}}\,\frac{1}{g_i}\,.
\end{gather}
The above relations are not well defined for all signs of the components in $\cG$, so one
needs to make a choice, which we take to be $g^0<0$ and $g_i>0$ throughout the discussion
below.

In order to obtain the attractor solution for given charges and the full flow to the asymptotic
vacuum described by \eqref{eq:G-STU}-\eqref{eq:vac-STU}, one needs to solve \eqref{eq:Gam-BB}
for the vector $\cB$. For completeness, we take the components of $\cB$ as
\begin{equation}
\cB = \left( \beta^0\,, \quad \beta^i \,, \quad \beta_i \,, \quad \beta_0 \right){}^{\!T}\,,
\end{equation}
and give explicitly the left hand side of \eqref{B-sol-fin}, as
\begin{equation}\label{eq:I4BBG-ex}
\tfrac14\,I^\prime_4(\cB,\cB,\cG) = 
\begin{pmatrix}
-g^0 \, \beta^I\beta_I - \beta^0\, g^I\beta_I
\\ 
\\ 
- \beta^i\, g^I\beta_I   + 2\,\beta^i \sum_{ j \neq i}\beta^jg_j 
 - c^{ijk}( 2\,\beta^0 \beta_j g_k + g_0 \beta_j \beta_k )
\\ 
\\ 
g_i\, \beta^I\beta_I + \beta_i\, g^I\beta_I  - 2\,\beta_i \sum_{ j \neq i}\beta^jg_j 
 - 2\,g_i \sum_{ j \neq i}\beta^j\beta_j 
\\ 
\\ 
+ \beta_0\, g^I\beta_I  + c^{ijk}\,g_i \beta_j \beta_k 
\end{pmatrix}\,,
\end{equation}
where we used the shorthand notation
\begin{align}
\beta^I\beta_I = \beta^0 \beta_0 + \beta^i \beta_i\,,
\qquad
g^I\beta_I  = g^0 \beta_0 + g_i \beta^i \,,
\end{align}
and $c^{ijk}$ is a completely symmetric tensor which is $c^{123}=1$ and and vanishes
when any two indices are equal.

It is useful to note that the rescaling
\begin{gather}
\beta^0\rightarrow \frac{g_0}{\sqrt{I_4(G)}} \,\hat\beta^0\,, \qquad
\beta^i\rightarrow \frac{\hat\beta^i}{g_i}\,, \qquad
\beta_i\rightarrow \frac{g_i}{\sqrt{I_4(G)}}\,\hat\beta_i\,, \qquad
\beta_0\rightarrow \frac{\hat\beta_0}{g^0}\,, \qquad
\nonumber\\
p^0\rightarrow \frac{g_0}{\sqrt{I_4(G)}} \,\hat p^0\,, \qquad
p^i\rightarrow \frac{\hat p^i}{g_i}\,, \qquad
q_i\rightarrow \frac{g_i}{\sqrt{I_4(G)}}\,\hat q_i\,, \qquad
q_0\rightarrow \frac{\hat q_0}{g^0}\,, \qquad
\label{eq:resc-g}
\end{gather}
on the components of $\cB$ and the charges, eliminates all explicit gaugings from
\eqref{B-sol-fin} when \eqref{eq:I4BBG-ex} is used. One can therefore solve \eqref{B-sol-fin}
with $I^\prime_4(\cB,\cB,\cG)$ as in \eqref{eq:I4BBG-ex} with all $g^0=g_i=1$ and re-introduce
them at the end using the inverse of \eqref{eq:resc-g} for both the components of $\cB$ and
the charges. Note that this is only relevant when the gauging is in the frame \eqref{eq:G-STU},
or frames obtained from it by simple enough dualities. One such example is the duality
\begin{equation}
p^0 \rightarrow p^0\,, \qquad p^i \rightarrow q_i\,, \qquad q_i \rightarrow - p^i\,, \qquad q_0\rightarrow q_0\,,
\end{equation}
which leads to a prepotential, $F\sim \sqrt{X^0 X^1 X^2 X^3}$, that can be uplifted to M-theory,
provided all gauging parameters are equal \cite{Cvetic:1999xp, Klemm:2011xw}. It follows that
the expressions given in terms of the rescaled quantities \eqref{eq:resc-g} can be directly used
for such an uplift, up to a possible convention dependent redefinition of the parameters.
Similar (but more complicated) rescalings can be obtained in all other frames by
duality, but we are not aware of any physical meaning associated to this operation.

\subsection{The $t^3$ model}
The first model we consider is the $t^3$ model, defined by the prepotential
\begin{equation}\label{t3}
 F= \frac{(X^1)^3}{X^0}\,,
\end{equation}
and is very useful in building intuition for the general case, since all symmetric
models can be truncated down to it. Indeed, we will view the $t^3$ model as the
special case of the STU model for which all three scalars are equal. We therefore
use the various relations \eqref{eq:G-STU}-\eqref{eq:I4BBG-ex} with all indices
$1$, $2$, $3$ equal\footnote{Note that this is equivalent, but not exactly identical
to the $t^3$ model one would find directly from \eqref{t3}, since for example we
use a modified inner product for which e.g. $\Iprod{\Gamma}{G}=g^0q_0-3\,g_1p^1$,
arising from the identification of the three scalars in the STU model. These
differences can be undone by appropriate rescalings, which we ignore for simplicity.}.

We now consider the most general vectors $\Gamma$ and $\cB$ allowed by \eqref{A-comm}
and \eqref{Dir-quant}, given by
\begin{align}\label{eq:GCh-t3}
\cB = &\, \left( -\frac{3\, g^0}{g_1}\, \frac{g^0\beta_0 + g_1\beta^1}{\kappa+4\,g_1p^1}\,q_1\,, \quad 
            \beta^1 \,, \quad           \frac{g^0\beta_0 + g_1\beta^1}{\kappa+4\,g_1p^1}\,q_1 \,, \quad
      \beta_0 \right){}^{\!T}\,,
\nonumber\\
\Gamma =&\, \left( -\frac{3\, g^0}{g_1}\, q_1\,,\quad p^1\,,\quad q_1\,,\quad \frac{1}{g^0}\,(\kappa+3\,g_1p^1) \right){}^{\!T}
\,,
\end{align}
where we note that the two parameters $\beta^1$, $\beta_0$ are constrained by \eqref{eq:B-constr},
so that there is only one independent component in $\cB$. It follows that \eqref{B-sol-fin} can
only be solved if the two charges $p^1$, $q_1$ are related, so that the general solution is
parametrised by a single charge. One finds that there are two branches of solutions, which we now
discuss in turn.

The simpler of the two branches arises by setting $q_1=0$, so that there is only a $p^1$ charge, that
we take as independent, and a $q_0$ charge which we take to be fixed as in \eqref{eq:GCh-t3}. The two
parameters, $\beta^1$, $\beta_0$ read
\begin{align}\label{eq:t3-p}
\beta_0 =&\, -\frac{1}{4\, g^0}\, \left(3\, \sqrt{\kappa + 4\, g_1 p^1 } + \sqrt{\kappa + 12\, g_1 p^1 } \right)\,,
\nonumber\\
\beta^1 =&\, \frac{1}{4\, g_1}\, \left( \sqrt{\kappa + 12\, g_1 p^1 } -\sqrt{\kappa + 4\, g_1 p^1 }  \right)\,,
\end{align}
which reproduce the result of \cite{Cacciatori:2009iz, Hristov:2010ri, Dall'Agata:2010gj}, up to the shift
\eqref{eq:B-shift}. Note that there is a lower limit for $p^1$ for this solution to be regular, i.e.
$\kappa + 4\, g_1 p^1 >0$. In this configuration, the axion is trivial throughout the flow.

The second branch allows for all charges to be nonzero, though fixed in terms of a single independent
parameter. We find that the charge vector is as in \eqref{eq:GCh-t3} with the additional constraint
\begin{equation}\label{eq:q-t3}
q_1 = \pm\frac{\sqrt{3}}{4\sqrt{-g_0 g^1}}\,(\kappa + 4\, g_1 p^1)\,,
\end{equation}
while the solution for the $\beta^1$, $\beta_0$ takes the form
\begin{align}\label{eq:t3-pq}
\beta_0 =&\, -\frac{1}{4\,\sqrt{2}\, g^0}\, \left( 4\,\sqrt{-(\kappa + 3\, g_1 p^1)} - 3\, \sqrt{-(\kappa + 4\, g_1 p^1) }\right)\,,
\nonumber\\
\beta^1 =&\, \frac{1}{4\,\sqrt{2}\, g_1}\, \left( 4\,\sqrt{-(\kappa + 3\, g_1 p^1)} +\sqrt{-(\kappa + 4\, g_1 p^1) }  \right)\,,
\end{align}
for both signs in \eqref{eq:q-t3}.
This is similar to \eqref{eq:t3-p} but has instead the opposite behaviour with respect to the magnetic
charge that is now subject to an upper bound, $(\kappa + 3\, g_1 p^1)<0$, by regularity. Indeed, one
may not turn off the electric charge and the axion continuously, unless $\kappa=-1$, which corresponds
to solutions with a hyperbolic horizon.

We have checked explicitly that both \eqref{eq:t3-p} and \eqref{eq:t3-pq} lead to physical solutions that
satisfy all constraints considered above, in the corresponding domain for the charge $p^1$. In particular,
they both lead to a finite area horizon consistent with \eqref{eq:entr-exp} and respect \eqref{BG-bound}.
We refrain from giving the full expressions for the scalar, since they are not particularly illuminating,
especially in the case of the solution with a running axion \eqref{eq:t3-pq}. However, we do note the
interesting fact that the quartic invariant of the physical charge in \eqref{eq:GCh-t3} is always positive
for both solutions in \eqref{eq:t3-p} and \eqref{eq:t3-pq}.

It is perhaps surprising to find this relatively rich set of BPS solutions, given that there is only a
single complex scalar. It would be very interesting to obtain a holographic interpretation of the two
solutions above, especially for the case including the axion. Presumably, one may interpret this as a
broken phase controlled by an additional VEV.

\subsection{The STU model}
We now turn to the more general STU model, defined by the prepotential \eqref{STU} which is routinely
used as a benchmark example for all symmetric models in the $SO(2,n+1)$ series. Using intuition from the
$t^3$ model, we can in fact solve \eqref{eq:G-STU}-\eqref{eq:I4BBG-ex} in the case of generic charges.
The result is a solution with five independent charges, which realises the two parameter duality boost
described recently in \cite{Halmagyi:2013uza}. Given the complexity of the general equations, we will use
the rescaled variables \eqref{eq:resc-g} throughout the discussion, suppressing the explicit hat. 

We first give the analogue of the solution \eqref{eq:t3-p} in the STU model, i.e. a solution with charge
\begin{gather}\label{eq:GCh-stu-p}
\Gamma = \left( 0 \,,\quad p^i\,,\quad 0\,,\quad \kappa + \sum p^i \right){}^{\!T}\,,
\nonumber\\
\cB = \left( 0\,, \quad \beta^i \,, \quad 0 \,, \quad \beta_0 \right){}^{\!T}\,,
\end{gather}
for reasons of comparison with \cite{Cacciatori:2009iz, Gnecchi:2013mta}, where the
same solution was given, but with a vector, $\cB$, shifted as in \eqref{eq:B-shift}, relative to the one
used here. Solving \eqref{B-sol-fin} with the explicit expression \eqref{eq:I4BBG-ex}, we obtain the
components
\begin{align}
\beta^i= &\, \frac14\, \left( 2\,\sqrt{\Delta^i} - \sum_j \sqrt{\Delta^j} + \sqrt{\sum_j \Delta^j - 2\, \kappa} \right) \,, 
\nonumber\\ 
\beta_0 = &\, -\frac14\, \left( \sum_i \sqrt{\Delta^i} + \sqrt{\sum_i \Delta^i - 2\, \kappa} \right)\,,
\end{align}
where we use the quantity
\begin{equation}
\Delta^1 = \frac{(\kappa + 2\, p^1 + 2\, p^3)(\kappa + 2\, p^1 + 2\, p^2)}{(\kappa + 2\, p^2 + 2\, p^3)}\,,
\end{equation}
and its cyclic permutations.

We now turn to solutions including electric charges and axions, so that we consider the following vectors
\begin{gather}\label{eq:GCh-stu-pq}
\Gamma = \left( -\sum q_i \,,\quad p^i\,,\quad q_i\,,\quad \kappa + \sum p^i \right){}^{\!T}
\nonumber\\
\cB = \left( -\sum \beta_i\,, \quad \beta^i \,, \quad \beta_i \,, \quad \beta_0 \right){}^{\!T}\,,
\,,
\end{gather}
which satisfy three of the conditions \eqref{A-comm}-\eqref{Dir-quant}, but do not mutually commute. We did
not impose this condition yet, as it seems to complicate the equations in \eqref{B-sol-fin} in first instance.
However, the condition $\Iprod{\Gamma}{\cB}=0$ imposes a constraint on the charges of the resulting solution,
which turns out to be given in terms of the parameter
\begin{equation}\label{eq:del-def}
\delta_1 = \pm \frac{(q_1 + q_2)(q_1 + q_3)}{(\kappa + 2\, p^1 + 2\, p^2)\,(\kappa + 2\, p^1 + 2\, p^3)}\,,
\end{equation}
and its cyclic permutations, $\delta_2$, $\delta_3$, as
\begin{equation}
  \frac1{\delta_1} + \frac1{\delta_2} + \frac1{\delta_3} = 1\,.
\end{equation}
Note that the two signs in the definition of the parameters $\delta_i$ in \eqref{eq:del-def} lead to two
branches of solutions. These relations are analogous to \eqref{eq:q-t3} and reduce to it in upon truncation
to the $t^3$ model.

The electrically charged solutions are given in terms of the quantity
\begin{equation}
\Delta^1 = \sqrt{\frac{(1-\delta_1)}{(1-\delta_2)(1-\delta_3)}
           \frac{(\kappa + 2\, p^1 + 2\, p^3)(\kappa + 2\, p^1 + 2\, p^2)}{(\kappa + 2\, p^2 + 2\, p^3)}}\,,
\end{equation}
and its cyclic permutations. The components of the vector $\cB$ are then given by
\begin{align}
 \beta_1 = &\,
\frac{-(q_2 + q_3)\, \Delta^1 + (q_1 + q_3)\, \Delta^2 + (q_1 + q_2)\, \Delta^3 }
{4\, {\Delta^1\Delta^2\Delta^3}}\, \sum_i  {\Delta^i}
- \frac{(2\,q_1 + q_3 + q_2)}{2\, {\Delta^1}}\,,
\nonumber\\
\beta^i= &\, \frac14\, \left( 2\, {\Delta^i} - \sum_i  {\Delta^i} +  {\Delta^0} \right)
\nonumber\\ 
\beta_0 = &\, -\frac14\, \left( \sum_i  {\Delta^i} + \sqrt{\Delta^0} \right)\,,
\nonumber\\ 
\Delta^0 = &\, 16\,\left( \sum_i \Delta^i \right)^2 
         - 4\,\left( \beta_1 \beta_2 + \beta_1 \beta_3 + \beta_2 \beta_3 \right) 
             - 8\, \left( \kappa + p^1 + p^2 + p^3 \right)\,,
\end{align}
where $\beta_2$ and $\beta_3$ are given by the obvious cyclic permutations of $\beta_1$.
Once again, the truncation to the $t^3$ model leads to the expression \eqref{eq:t3-pq} and we find that
all the requirements for a physical solution are satisfied if the $p^i<0$. We refrain from giving further
explicit expressions for the scalars, as they are not particularly illuminating.

\section{Conclusion and outlook}
\label{sec:conc}

In this paper, we have considered the general static $1/4$-BPS flow in Fayet-Iliopoulos gauged $\cN\!=\!2$
supergravity coupled to vector multiplets describing a symmetric scalar manifold. Using techniques from
very special geometry, we have shown that it is possible to reduce the BPS equations to a form where the
scalar fields only appear through a single combination that transforms covariantly under electric/magnetic
duality reparametrisations. Considering a general expansion at the horizon, we have found that physical
black hole solutions are very restricted, their basic properties being captured by the standard ansatz
that was proposed for the simplified axion-free case
\cite{Cacciatori:2009iz,Dall'Agata:2010gj, Hristov:2010ri, Gnecchi:2013mta, Halmagyi:2013uza}.

In addition, we have found that the full flow is parametrised in terms of a real symplectic vector that
is determined by the gauging and charge vectors and can moreover be identified with the solution to the
attractor equations. The final result is a set of algebraic equations at the attractor, whose solution
implies the existence of a full asymptotically AdS$_4$ black hole solution. These equations involve the
quartic invariant of very special geometry taking as arguments both the charges and gaugings, in a way
similar to the two-charge invariants of \cite{Ferrara:2010ug, Andrianopoli:2011gy, Ceresole:2011xd}. We
have given a number of explicit examples of solutions to these equations, describing black holes with all
allowed electric and magnetic charges turned on, as well as nontrivial axions, in the STU model and its
truncations.

In view of the relative complexity of the original flow equations, this is a somewhat surprising result.
For example, the K\"ahler connection appears to be necessarily trivial, despite the fact that the BPS
equations do not impose that requirement directly. Instead, demanding regularity of black hole solutions
one finds that the corresponding BPS equation must be trivially satisfied, with the terms involved vanishing
individually\footnote{This particular simplification is however an artifact of a restriction that can
be relaxed, see the notes at the end of sections \ref{sec:attr} and \ref{sec:flow-ana}.}. However, we
expect this to change in the case of rotating and/or NUT-charged black holes
\cite{Klemm:2011xw,Colleoni:2012jq}. Indeed, this was shown to be true for the asymptotically flat
solutions to the theory in \cite{Hristov:2012nu}, where the scalar flow of the asymptotically flat extremal
non-BPS black holes \cite{Bossard:2013oga} was shown to be very similar to the
one studied in this paper.

It is important to note that the method used to simplify the BPS flow equations in this paper is
straightforward to apply in more general situations, for example in general stationary solutions and in
theories involving hypermultiplets, as long as the vector multiplet scalars are parametrising a
symmetric manifold. The latter observation is particularly important, given that, in a general gauged
theory, the BPS flow for the vector multiplet scalars depends on the hyperscalars only through the moment
maps, arranged in a symplectic vector, $(P^I,\, P_I){}^{T}$, which generalises the vector of FI parameters,
$\cG$, considered in this paper. Since we did not make any use of the fact that $\cG$ is constant in
the derivation of the simplified flow equation, it follows that one may repeat exactly the same steps to
simplify the general BPS conditions. Finally, it is interesting to consider the possibility of extending
the ansatze proposed for axion-free non-BPS solutions to FI gauged supergravity
\cite{Klemm:2012yg,Toldo:2012ec,Klemm:2012vm,Gnecchi:2012kb} to the general case, using the same method.
While the systems mentioned above are considerably more complicated, we expect that the steps followed for
the system studied in this paper will still be be useful.
We hope to return to some of these issues in the future.

\section*{Acknowledgement}
The author wishes to thank Kiril Hristov for introducing him to this subject through many valuable
discussions and for comments on an earlier version of this paper. He further thanks Nick Halmagyi for
discussions and critical comments on an earlier version of this paper.
Further useful discussions with Guillaume Bossard and Alessandro Tomasiello are gratefully acknowledged.
The work of S.K. is supported by the European Research Council under the European Union's Seventh
Framework Program (FP/2007-2013)-ERC Grant Agreement n. 307286 (XD-STRING). His research is also
supported in part by INFN.

\begin{appendix}

\section[Conventions on N=2 supergravity]{Conventions on $\cN\!=\!2$ supergravity}
\label{app:conv}

In this paper we follow the notation and conventions of \cite{Bossard:2012xsa}.
In this appendix we collect some basic definitions that are useful in the main
text, referring to that paper for more details.

The vector fields naturally arrange in a symplectic vector of electric and magnetic
gauge field strengths, whose integral over a sphere defines the associated
electromagnetic charges as
\begin{equation}\label{eq:dual-gauge}
 \cF_{\mu\nu}=\begin{pmatrix} F_{\mu\nu}^I\\ G_{I\, \mu\nu}\end{pmatrix}\,,
\qquad
\Gamma=\begin{pmatrix} p^I\\ q_{I}\end{pmatrix}
 =\frac{1}{2\pi} \,\int_{S^2} \cF\,.
\end{equation}
Here, $F_{\mu\nu}^I$ are the field strengths of the vector fields, while the $G_{I\, \mu\nu}$
stand for the dual field strengths defined by taking a derivative of the Lagrangian or,
equivalently, by the scalar dependent period matrix, $\cN_{IJ}$, as
\begin{equation}\label{G-def}
  G^-_{\mu\nu}{}_I = \cN_{IJ} F^-_{\mu\nu}{}^J\,,
 \end{equation}
where the explicit form of the period matrix will not be used.
 
The physical scalar fields $t^i$, which parametrize a special K\"ahler space of complex dimension $\nv$, appear through the so called symplectic section, $\cV$.
Choosing a basis, this section can be written in components in terms of scalars
$X^I$ as
\begin{equation}\label{eq:sym-sec}
\cV=\begin{pmatrix} X^I\\ F_I\end{pmatrix}\,, \qquad
F_I= \frac{\partial F}{\partial X^I}\,,
\end{equation}
where $F$ is a holomorphic function of degree two, called the prepotential,
which we will always consider to be cubic
\begin{equation}\label{prep-def}
F=-\frac{1}{6}c_{ijk}\frac{X^i X^j X^k}{X^0} \,,
\end{equation}
for completely symmetric $c_{ijk}$, $i=1,\dots \nv$. The section $\mathcal{V}$ is subject to the constraints
\begin{equation}
  \label{eq:D-gauge}
  \Iprod{\bar{\mathcal{V}}}{\mathcal{V}} =  i \qquad
  \Iprod{\bar{D}_{\bar i}\bar{\mathcal{V}}}{D_j\mathcal{V}} =  -i\,g_{\bar{i}j} \,,
\end{equation}
with all other inner products vanishing, and is uniquely determined by the physical scalar fields $t^i=\frac{X^i}{X^0}$ up to a local $U(1)$ transformation. Here, $g_{\bar{\imath}j}$ is the K\"ahler metric and the K\"ahler covariant derivative $D_i\cV$ contains the K\"{a}hler connection $Q_\mu$, defined through the K\"{a}hler potential as
\begin{equation} Q = \I[ \partial_ i \cK\, dt^i] \label{Kah-def}\ ,\qquad
 \cK = - \mbox{ln}\left( \tfrac{i}6\, c_{ijk} (t-\bar t)^i(t-\bar t)^j(t-\bar t)^k\right) \,. \end{equation}

We introduce the following notation for any symplectic vector $\Gamma$
\begin{align}\label{ch-def}
Z(\Gamma) = \Iprod{\Gamma}{\cV} \,,\qquad
Z_i(\Gamma) = \Iprod{\Gamma}{D_i \cV} \,,
\end{align}
with the understanding that when an argument does not appear explicitly, the
vector of charges in \eqref{eq:dual-gauge} should be inserted. In addition,
when the argument is form valued, the operation is applied component wise.
With these definitions it is possible to introduce a scalar dependent complex
basis for symplectic vectors, given by $(\cV,\, D_i\cV)$, so that any vector
$\Gamma$ can be expanded as
\begin{equation}\label{Z-expand}
\Gamma = 2 \I[- \bar{Z}(\Gamma)\,\cV + g^{\bar \imath j} \bar{Z}_{\bar \imath}(\Gamma)\, D_j \cV]\,,
\end{equation}
whereas the symplectic inner product can be expressed as
\begin{equation}\label{inter-prod-Z}
\Iprod{\Gamma_1}{\Gamma_2} = 2 \I[- Z(\Gamma_1)\,\bar{Z}(\Gamma_2)
   +g^{i\bar \jmath} Z_i(\Gamma_1) \, \bar{Z}_{\bar \jmath}(\Gamma_2)]\,.
\end{equation}
In addition, we introduce the scalar dependent complex structure $\mathrm{J}$,
defined as
\begin{equation}\label{CY-hodge}
\mathrm{J}\cV=-i \cV\,,\quad
\mathrm{J} D_i\cV=i  D_i\cV\,,
\end{equation}
which can be solved to determine $\mathrm{J}$ in terms of the period matrix $\cN_{IJ}$
in \eqref{G-def}, see e.g.~\cite{Ceresole:1995ca} for more details.
With this definition, we can express the complex self-duality of the gauge field strengths as
\begin{equation}
 \mathrm{J}\,\cF=-*\cF\,,\label{cmplx-sdual}
\end{equation}
which is the duality covariant form of the relation between electric and magnetic
components. Finally, we record the important relation
\begin{equation}\label{VBH-def}
 \tfrac12\,\Iprod{\Gamma}{\mathrm{J}\,\Gamma}=
 |Z(\Gamma)|^2 + g^{i\bar \jmath} Z_i(\Gamma) \, \bar{Z}_{\bar \jmath}(\Gamma)
 \equiv V_{\text{\tiny BH}}(\Gamma)\,,
\end{equation}
where we defined the black hole potential $V_{\text{\tiny BH}}(\Gamma)$.

\section{Identities involving the quartic invariant}
\label{app:I4}
In this short appendix, we summarise a number of useful relations involving the
quartic invariant, defined for all symmetric models. The starting point is the
definition of the invariant, $I_4(\Gamma)$, for any symplectic vector, $\Gamma$,
as \cite{Ferrara:1997uz,Ferrara:2006yb}
\begin{eqnarray}
I_4(\Gamma)&=& \frac{1}{4!} t^{MNPQ}\Gamma_M\Gamma_N\Gamma_P\Gamma_Q
\nonumber\\
         &=& - (p^0 q_0 + p^i q_i)^2 + \frac{2}{3} \,q_0\,c_{ijk} p^i p^j p^k- \frac{2}{3} \,p^0\,c^{ijk} q_i q_j q_k 
             + c_{ijk}p^jp^k\,c^{ilm}q_lq_m\,, \label{I4-ch}
\end{eqnarray}
where $M$, $N$\dots are indices encompassing both electric and magnetic components
and we also defined the completely symmetric tensor $t^{MNPQ}$ for later reference.
It is also convenient to define a symplectic vector out the first derivative, $I_4^\prime(\Gamma)$,
of the quartic invariant, as
\begin{equation} \label{I4-der-basis}
I_4^{\prime}(\Gamma)_{M} \equiv \Omega_{MN}\frac{\partial I_4(\Gamma)}{\partial \Gamma_N} 
 = \frac{1}{3!}\, \Omega_{MN} t^{NPQR} \Gamma_P \Gamma_Q \Gamma_R \,,
\end{equation}
where $\Omega^{MN}$ is the symplectic form, so that the following relations hold
\begin{equation}
 \Iprod{\Gamma}{I^\prime_4(\Gamma)} = 4\,I_4(\Gamma)  \ , \qquad I^\prime_4(\Gamma,\Gamma,\Gamma) = 6 I_4^\prime(\Gamma) \ .
\end{equation}
Throughout this paper, all instances of $I_4(\Gamma_1,\Gamma_2,\Gamma_3,\Gamma_4)$ will denote
the contraction of the tensor $t^{MNPQ}$ in \eqref{I4-ch} with the four charges, without any symmetry
factors, except for the case with a single argument, as in $I_4(\Gamma)$ and $I^\prime_4(\Gamma)$.
For more details on this tensor, see \cite{Bossard:2013oga} in the real basis and
\cite{Ferrara:2011di} in the complex basis, to be defined shortly.

We now record some identities that are used repeatedly in the main text, starting with the
fundamental property
\begin{equation}
 I_4^{\prime}( I_4^{\prime}(\Gamma) ) = -16\, I_4(\Gamma)^2 \Gamma\,.
\end{equation}
Further properties include the septic and quintic identities 
\begin{equation}\label{eq:quint}
 I_4^{\prime}( I_4^{\prime}(\Gamma), I_4^{\prime}(\Gamma), \Gamma ) = 8\, I_4(\Gamma)\, I_4^{\prime}(\Gamma)\,,
 \qquad
 I_4^{\prime}( I_4^{\prime}(\Gamma), \Gamma, \Gamma ) = -8\, I_4(\Gamma) \Gamma \,.
\end{equation}
Finally, the projection operator
\begin{equation}\label{eq:proj}
 I_4^{\prime}( I_4^{\prime}(\Gamma), \Gamma, \cB ) 
 = 2\, \Iprod{\Gamma}{\cB}\,I_4^{\prime}(\Gamma) + 2\, \Iprod{I_4^{\prime}(\Gamma)}{\cB}\,\Gamma \,,
\end{equation}
where $\cB$ is an arbitrary vector, is particularly useful in the analysis of
sections \ref{sec:attr} and \ref{sec:AdS}.

One can rewrite the quartic invariant in the complex basis \cite{Cerchiai:2009pi}, leading
to the following alternative definition
\begin{align} 
I_4(\Gamma) =&\, \left( Z \, \bar Z - Z_i \, \bar Z^i \right)^2 
- c_{mij} \bar Z^i \bar Z^j \, c^{mkl} Z_k  Z_l 
+  \frac{2}{3} \, \bar Z \, c^{ijk} Z^i Z^j Z^k  + \frac{2}{3} \, Z \,c_{ijk} \bar Z^i \bar Z^j \bar Z^k\,.
\label{I4-def}
\end{align}
Despite the appearance of the central charges, this expression is by construction
independent of the scalars, which only appear due to the change of basis in \eqref{Z-expand}.
The derivatives of \eqref{I4-def} with respect to the central charges $Z(\Gamma)$ and
$Z_i(\Gamma)$ can be used to define the tensor $t^{MNPQ}$ and its contractions in the
complex basis, in exactly the same way as above. We will not make use of this basis,
but we do note an identity central to the analysis of sections \ref{sec:attr} and
\ref{sec:AdS}. Consider the contraction of \eqref{I4-def} with two instances of a charge
and two instances of the symplectic section itself, so that the resulting expression
is at most quadratic in the central charges $Z(\Gamma)$ and $Z_i(\Gamma)$, due to
\eqref{eq:D-gauge}. The resulting equality reads
\begin{eqnarray}
 \tfrac14\, I_4(\Gamma,\Gamma, 2 \I\cV, 2 \I\cV)
    &=& 4\,\left( |Z(\Gamma)|^2+ \R (Z(\Gamma))^2 \right)
    -2\,(|Z(\Gamma)|^2+ Z_i(\Gamma)\bar Z(\Gamma)^i)
      \,,
    \label{I4toVbh}
\end{eqnarray}
while its derivative is
\begin{eqnarray}
 \tfrac12\, I_4(\Gamma, 2 \I\cV, 2 \I\cV)
    &=& 8\,\I(Z(\Gamma))\,\I\cV
    +16\,\R(Z(\Gamma))\,\R\cV
    -2\,\mathrm{J}\,\Gamma
      \,.
    \label{I4toJ}
\end{eqnarray}
The last equation relates the action of the scalar dependent complex structure on the
charge to a matrix operation involving the quartic invariant. Since the quartic invariant
is evaluated with two instances of the symplectic section, this form is particularly
useful in solving the BPS equations for black hole backgrounds, where an ansatz for the
imaginary part of the section is usually considered.

\end{appendix}

\bibliographystyle{utphys}
\bibliography{AdSaxions}

\providecommand{\href}[2]{#2}\begingroup\raggedright\begin{thebibliography}{10}

\bibitem{Behrndt:1997ny}
K.~Behrndt, D.~Lust, and W.~A. Sabra, ``{Stationary solutions of N = 2
  supergravity},'' \href{http://dx.doi.org/10.1016/S0550-3213(97)00633-0}{{\em
  Nucl. Phys.} {\bfseries B510} (1998) 264--288},
\href{http://arxiv.org/abs/hep-th/9705169}{{\ttfamily arXiv:hep-th/9705169}}.

\bibitem{Denef:2000nb}
F.~Denef, ``{Supergravity flows and D-brane stability},''
  \href{http://dx.doi.org/10.1088/1126-6708/2000/08/050}{{\em JHEP} {\bfseries
  0008} (2000) 050},
\href{http://arxiv.org/abs/hep-th/0005049}{{\ttfamily arXiv:hep-th/0005049
  [hep-th]}}.

\bibitem{Gauntlett:2002nw}
J.~P. Gauntlett, J.~B. Gutowski, C.~M. Hull, S.~Pakis, and H.~S. Reall, ``{All
  supersymmetric solutions of minimal supergravity in five dimensions},''
  \href{http://dx.doi.org/10.1088/0264-9381/20/21/005}{{\em Class. Quant.
  Grav.} {\bfseries 20} (2003) 4587--4634},
\href{http://arxiv.org/abs/hep-th/0209114}{{\ttfamily arXiv:hep-th/0209114}}.

\bibitem{Gauntlett:2004qy}
J.~P. Gauntlett and J.~B. Gutowski, ``{General concentric black rings},''
  \href{http://dx.doi.org/10.1103/PhysRevD.71.045002}{{\em Phys.Rev.}
  {\bfseries D71} (2005) 045002},
\href{http://arxiv.org/abs/hep-th/0408122}{{\ttfamily arXiv:hep-th/0408122
  [hep-th]}}.

\bibitem{Cacciatori:2008ek}
S.~L. Cacciatori, D.~Klemm, D.~S. Mansi, and E.~Zorzan, ``{All timelike
  supersymmetric solutions of N=2, D=4 gauged supergravity coupled to abelian
  vector multiplets},''
  \href{http://dx.doi.org/10.1088/1126-6708/2008/05/097}{{\em JHEP} {\bfseries
  05} (2008) 097},
\href{http://arxiv.org/abs/0804.0009}{{\ttfamily arXiv:0804.0009 [hep-th]}}.

\bibitem{Klemm:2010mc}
D.~Klemm and E.~Zorzan, ``{The timelike half-supersymmetric backgrounds of N=2,
  D=4 supergravity with Fayet-Iliopoulos gauging},''
  \href{http://dx.doi.org/10.1103/PhysRevD.82.045012}{{\em Phys.Rev.}
  {\bfseries D82} (2010) 045012},
\href{http://arxiv.org/abs/1003.2974}{{\ttfamily arXiv:1003.2974 [hep-th]}}.

\bibitem{Meessen:2012sr}
P.~Meessen and T.~Ortin, ``{Supersymmetric solutions to gauged N=2 d=4 sugra:
  the full timelike shebang},''
  \href{http://dx.doi.org/10.1016/j.nuclphysb.2012.05.023}{{\em Nucl.Phys.}
  {\bfseries B863} (2012) 65--89},
\href{http://arxiv.org/abs/1204.0493}{{\ttfamily arXiv:1204.0493 [hep-th]}}.

\bibitem{Cacciatori:2009iz}
S.~L. Cacciatori and D.~Klemm, ``{Supersymmetric AdS(4) black holes and
  attractors},'' \href{http://dx.doi.org/10.1007/JHEP01(2010)085}{{\em JHEP}
  {\bfseries 1001} (2010) 085},
\href{http://arxiv.org/abs/0911.4926}{{\ttfamily arXiv:0911.4926 [hep-th]}}.

\bibitem{Dall'Agata:2010gj}
G.~Dall'Agata and A.~Gnecchi, ``{Flow equations and attractors for black holes
  in N = 2 U(1) gauged supergravity},''
  \href{http://dx.doi.org/10.1007/JHEP03(2011)037}{{\em JHEP} {\bfseries 03}
  (2011) 037},
\href{http://arxiv.org/abs/1012.3756}{{\ttfamily arXiv:1012.3756 [hep-th]}}.

\bibitem{Hristov:2010ri}
K.~Hristov and S.~Vandoren, ``{Static supersymmetric black holes in $AdS_4$
  with spherical symmetry},''
  \href{http://dx.doi.org/10.1007/JHEP04(2011)047}{{\em JHEP} {\bfseries 04}
  (2011) 047},
\href{http://arxiv.org/abs/1012.4314}{{\ttfamily arXiv:1012.4314 [hep-th]}}.

\bibitem{Klemm:2011xw}
D.~Klemm, ``{Rotating BPS black holes in matter-coupled $AdS_4$
  supergravity},'' \href{http://dx.doi.org/10.1007/JHEP07(2011)019}{{\em JHEP}
  {\bfseries 1107} (2011) 019},
\href{http://arxiv.org/abs/1103.4699}{{\ttfamily arXiv:1103.4699 [hep-th]}}.

\bibitem{Barisch:2011ui}
S.~Barisch, G.~Lopes~Cardoso, M.~Haack, S.~Nampuri, and N.~A. Obers, ``{Nernst
  branes in gauged supergravity},''
  \href{http://dx.doi.org/10.1007/JHEP11(2011)090}{{\em JHEP} {\bfseries 1111}
  (2011) 090},
\href{http://arxiv.org/abs/1108.0296}{{\ttfamily arXiv:1108.0296 [hep-th]}}.

\bibitem{Colleoni:2012jq}
M.~Colleoni and D.~Klemm, ``{Nut-charged black holes in matter-coupled N=2, D=4
  gauged supergravity},''
  \href{http://dx.doi.org/10.1103/PhysRevD.85.126003}{{\em Phys.Rev.}
  {\bfseries D85} (2012) 126003},
\href{http://arxiv.org/abs/1203.6179}{{\ttfamily arXiv:1203.6179 [hep-th]}}.

\bibitem{Halmagyi:2013sla}
N.~Halmagyi, M.~Petrini, and A.~Zaffaroni, ``{BPS black holes in $AdS_{4}$ from
  M-theory},'' \href{http://dx.doi.org/10.1007/JHEP08(2013)124}{{\em JHEP}
  {\bfseries 1308} (2013) 124},
\href{http://arxiv.org/abs/1305.0730}{{\ttfamily arXiv:1305.0730 [hep-th]}}.

\bibitem{Halmagyi:2013qoa}
N.~Halmagyi, ``{BPS Black Hole Horizons in N=2 Gauged Supergravity},''
  \href{http://dx.doi.org/10.1007/JHEP02(2014)051}{{\em JHEP} {\bfseries 1402}
  (2014) 051},
\href{http://arxiv.org/abs/1308.1439}{{\ttfamily arXiv:1308.1439 [hep-th]}}.

\bibitem{Barisch-Dick:2013xga}
S.~Barisch-Dick, G.~L. Cardoso, M.~Haack, and A.~V\'{e}liz-Osorio, ``{Quantum
  corrections to extremal black brane solutions},''
  \href{http://dx.doi.org/10.1007/JHEP02(2014)105}{{\em JHEP} {\bfseries 1402}
  (2014) 105},
\href{http://arxiv.org/abs/1311.3136}{{\ttfamily arXiv:1311.3136 [hep-th]}}.

\bibitem{Gnecchi:2013mta}
A.~Gnecchi and N.~Halmagyi, ``{Supersymmetric Black Holes in AdS4 from Very
  Special Geometry},''
\href{http://arxiv.org/abs/1312.2766}{{\ttfamily arXiv:1312.2766 [hep-th]}}.

\bibitem{Halmagyi:2013uza}
N.~Halmagyi and T.~Vanel, ``{AdS Black Holes from Duality in Gauged
  Supergravity},'' \href{http://dx.doi.org/10.1007/JHEP04(2014)130}{{\em JHEP}
  {\bfseries 1404} (2014) 130},
\href{http://arxiv.org/abs/1312.5430}{{\ttfamily arXiv:1312.5430 [hep-th]}}.

\bibitem{Bellucci:2008cb}
S.~Bellucci, S.~Ferrara, A.~Marrani, and A.~Yeranyan, ``{d=4 Black Hole
  Attractors in N=2 Supergravity with Fayet-Iliopoulos Terms},''
  \href{http://dx.doi.org/10.1103/PhysRevD.77.085027}{{\em Phys.Rev.}
  {\bfseries D77} (2008) 085027},
\href{http://arxiv.org/abs/0802.0141}{{\ttfamily arXiv:0802.0141 [hep-th]}}.

\bibitem{Hristov:2012nu}
K.~Hristov, S.~Katmadas, and V.~Pozzoli, ``{Ungauging black holes and hidden
  supercharges},'' \href{http://dx.doi.org/10.1007/JHEP01(2013)110}{{\em JHEP}
  {\bfseries 1301} (2013) 110},
\href{http://arxiv.org/abs/1211.0035}{{\ttfamily arXiv:1211.0035 [hep-th]}}.

\bibitem{Halmagyi:2014qza}
N.~Halmagyi, ``{Static BPS Black Holes in AdS4 with General Dyonic Charges},''
\href{http://arxiv.org/abs/1408.2831}{{\ttfamily arXiv:1408.2831 [hep-th]}}.

\bibitem{Chow:2013gba}
D.~D.~K. Chow and G.~Comp\`{e}re, ``{Dyonic AdS black holes in maximal gauged
  supergravity},'' \href{http://dx.doi.org/10.1103/PhysRevD.89.065003}{{\em
  Phys.Rev.} {\bfseries D89} (2014) 065003},
\href{http://arxiv.org/abs/1311.1204}{{\ttfamily arXiv:1311.1204 [hep-th]}}.

\bibitem{Gnecchi:2013mja}
A.~Gnecchi, K.~Hristov, D.~Klemm, C.~Toldo, and O.~Vaughan, ``{Rotating black
  holes in 4d gauged supergravity},''
  \href{http://dx.doi.org/10.1007/JHEP01(2014)127}{{\em JHEP} {\bfseries 1401}
  (2014) 127},
\href{http://arxiv.org/abs/1311.1795}{{\ttfamily arXiv:1311.1795 [hep-th]}}.

\bibitem{Cvetic:1999xp}
M.~Cvetic, M.~Duff, P.~Hoxha, J.~T. Liu, H.~Lu, {\em et al.}, ``{Embedding AdS
  black holes in ten-dimensions and eleven-dimensions},''
  \href{http://dx.doi.org/10.1016/S0550-3213(99)00419-8}{{\em Nucl.Phys.}
  {\bfseries B558} (1999) 96--126},
\href{http://arxiv.org/abs/hep-th/9903214}{{\ttfamily arXiv:hep-th/9903214
  [hep-th]}}.

\bibitem{Ferrara:2010ug}
S.~Ferrara, A.~Marrani, E.~Orazi, R.~Stora, and A.~Yeranyan, ``{Two-Center
  Black Holes Duality-Invariants for stu Model and its lower-rank
  Descendants},'' \href{http://dx.doi.org/10.1063/1.3589319}{{\em J.Math.Phys.}
  {\bfseries 52} (2011) 062302},
\href{http://arxiv.org/abs/1011.5864}{{\ttfamily arXiv:1011.5864 [hep-th]}}.

\bibitem{Andrianopoli:2011gy}
L.~Andrianopoli, R.~D'Auria, S.~Ferrara, A.~Marrani, and M.~Trigiante,
  ``{Two-Centered Magical Charge Orbits},''
  \href{http://dx.doi.org/10.1007/JHEP04(2011)041}{{\em JHEP} {\bfseries 1104}
  (2011) 041},
\href{http://arxiv.org/abs/1101.3496}{{\ttfamily arXiv:1101.3496 [hep-th]}}.

\bibitem{Ceresole:2011xd}
A.~Ceresole, S.~Ferrara, A.~Marrani, and A.~Yeranyan, ``{Small Black Hole
  Constituents and Horizontal Symmetry},''
  \href{http://dx.doi.org/10.1007/JHEP06(2011)078}{{\em JHEP} {\bfseries 1106}
  (2011) 078},
\href{http://arxiv.org/abs/1104.4652}{{\ttfamily arXiv:1104.4652 [hep-th]}}.

\bibitem{Bossard:2013oga}
G.~Bossard and S.~Katmadas, ``{Duality covariant multi-centre black hole
  systems},'' \href{http://dx.doi.org/10.1007/JHEP08(2013)007}{{\em JHEP}
  {\bfseries 1308} (2013) 007},
\href{http://arxiv.org/abs/1304.6582}{{\ttfamily arXiv:1304.6582 [hep-th]}}.

\bibitem{Klemm:2012yg}
D.~Klemm and O.~Vaughan, ``{Nonextremal black holes in gauged supergravity and
  the real formulation of special geometry},''
  \href{http://dx.doi.org/10.1007/JHEP01(2013)053}{{\em JHEP} {\bfseries 1301}
  (2013) 053},
\href{http://arxiv.org/abs/1207.2679}{{\ttfamily arXiv:1207.2679 [hep-th]}}.

\bibitem{Toldo:2012ec}
C.~Toldo and S.~Vandoren, ``{Static nonextremal AdS4 black hole solutions},''
  \href{http://dx.doi.org/10.1007/JHEP09(2012)048}{{\em JHEP} {\bfseries 1209}
  (2012) 048},
\href{http://arxiv.org/abs/1207.3014}{{\ttfamily arXiv:1207.3014 [hep-th]}}.

\bibitem{Klemm:2012vm}
D.~Klemm and O.~Vaughan, ``{Nonextremal black holes in gauged supergravity and
  the real formulation of special geometry II},''
  \href{http://dx.doi.org/10.1088/0264-9381/30/6/065003}{{\em
  Class.Quant.Grav.} {\bfseries 30} (2013) 065003},
\href{http://arxiv.org/abs/1211.1618}{{\ttfamily arXiv:1211.1618 [hep-th]}}.

\bibitem{Gnecchi:2012kb}
A.~Gnecchi and C.~Toldo, ``{On the non-BPS first order flow in N=2 U(1)-gauged
  Supergravity},'' \href{http://dx.doi.org/10.1007/JHEP03(2013)088}{{\em JHEP}
  {\bfseries 1303} (2013) 088},
\href{http://arxiv.org/abs/1211.1966}{{\ttfamily arXiv:1211.1966 [hep-th]}}.

\bibitem{Bossard:2012xsa}
G.~Bossard and S.~Katmadas, ``{Duality covariant non-BPS first order
  systems},'' \href{http://dx.doi.org/10.1007/JHEP09(2012)100}{{\em JHEP}
  {\bfseries 1209} (2012) 100},
\href{http://arxiv.org/abs/1205.5461}{{\ttfamily arXiv:1205.5461 [hep-th]}}.

\bibitem{Ceresole:1995ca}
A.~Ceresole, R.~D'Auria, and S.~Ferrara, ``{The Symplectic structure of N=2
  supergravity and its central extension},''
  \href{http://dx.doi.org/10.1016/0920-5632(96)00008-4}{{\em
  Nucl.Phys.Proc.Suppl.} {\bfseries 46} (1996) 67--74},
\href{http://arxiv.org/abs/hep-th/9509160}{{\ttfamily arXiv:hep-th/9509160
  [hep-th]}}.

\bibitem{Ferrara:1997uz}
S.~Ferrara and M.~Gunaydin, ``{Orbits of exceptional groups, duality and BPS
  states in string theory},''
  \href{http://dx.doi.org/10.1142/S0217751X98000913}{{\em Int.J.Mod.Phys.}
  {\bfseries A13} (1998) 2075--2088},
\href{http://arxiv.org/abs/hep-th/9708025}{{\ttfamily arXiv:hep-th/9708025
  [hep-th]}}.

\bibitem{Ferrara:2006yb}
S.~Ferrara, E.~G. Gimon, and R.~Kallosh, ``{Magic supergravities, N= 8 and
  black hole composites},''
  \href{http://dx.doi.org/10.1103/PhysRevD.74.125018}{{\em Phys.Rev.}
  {\bfseries D74} (2006) 125018},
\href{http://arxiv.org/abs/hep-th/0606211}{{\ttfamily arXiv:hep-th/0606211
  [hep-th]}}.

\bibitem{Ferrara:2011di}
S.~Ferrara, A.~Marrani, and A.~Yeranyan, ``{On Invariant Structures of Black
  Hole Charges},'' \href{http://dx.doi.org/10.1007/JHEP02(2012)071}{{\em JHEP}
  {\bfseries 1202} (2012) 071},
\href{http://arxiv.org/abs/1110.4004}{{\ttfamily arXiv:1110.4004 [hep-th]}}.

\bibitem{Cerchiai:2009pi}
B.~L. Cerchiai, S.~Ferrara, A.~Marrani, and B.~Zumino, ``{Duality, Entropy and
  ADM Mass in Supergravity},''
  \href{http://dx.doi.org/10.1103/PhysRevD.79.125010}{{\em Phys.Rev.}
  {\bfseries D79} (2009) 125010},
\href{http://arxiv.org/abs/0902.3973}{{\ttfamily arXiv:0902.3973 [hep-th]}}.

\end{thebibliography}\endgroup
\end{document}